# Airbus A32x vs Boeing 737 Safety Occurrences

Graham Wild, *Member, IEEE*

*Abstract*—Safety is the priority for airlines. Airlines are sensitive to passengers' perceptions of safety, having randomly assigned the Boeing 737 Max to routes and times. Historically, Boeing has been considered more reliable and safer than Airbus. Hence, it is worth asking the question, are there any differences in the safety occurrences of the core narrow-body single-aisle aircraft of Boeing and Airbus; the 737 and A32x families of aircraft. Utilizing the International Civil Aviation Organization safety occurrence data, from 2008 to 2019, these aircraft were compared in terms of occurrence type, occurrence category, phase of flight, injury level, and fatalities. It was found that Boeing had more accidents than expected, while Airbus had less (p=0.015). In terms of fatalities Boeing has had more than expected, with Airbus less (p<0.001). Looking at just accidents, only the number of fatalities was statistically significantly different. In both cases, the increased number of fatalities for Boeing appears to be the result of the two recent Boeing 737 Max accidents. Looking at the reported fatal and hull loss accident rates, it was also found that the annual reduction for the Airbus A32x aircraft were better than for the Boeing 737 aircraft.

## I. INTRODUCTION

AIR transport globally, in terms of domestic and international travel, is slowly recovering post COVID-19. Traffic in 2022 was 68.5% of the revenue passenger kilometers (RPKs) reported in 2019, and up 64.4% compared to 2021 [1]. This suggests that it will take up to half a decade for the airline industry to fully recover, a similar time scale to the recovery after the September 11 terrorist attacks, and much longer than any intervening disruptions [2]. The future growth of the industry also needs to contend with net zero targets, for which the International Air Transport Association (IATA) has set a timeframe of 2050 [3]. This puts significant pressure on the airline industry which from 2000 to 2019 collectively averaged a profit margin of 1.03%, while the Standard and Poor's (S&P) 500 companies averaged 8.4% [4].

Despite consistent exponential traffic growth and significant financial pressures, the airline industry has an excellent safety record, and is second only to the Japanese Shinkansen (bullet train) in transport safety. The number of accidents per departure for passenger operations has reduced exponentially since 1945 [5], and for 2012 to 2020 the number of accidents has held constant at an average of 9.5 per year [6]. For this very reason the events of 2018 and 2019 involving the Boeing 737-MAX were considered out of character for the industry. These two events were the fatal accidents which killed 346 people [7]; specifically, the Lion Air accident on October 29, 2018, and the Ethiopian Airlines accident on March 10, 2019 [8-10]. Such events can be referred to as "black swans" [11], defined as "surprising extreme event(s) relative to the present knowledge"; however, in light of all the evidence, there are those that would not use such a label for the Boeing 737 MAX accidents [12].

Previous research looking at human factors related accidents considered the effect of manufacturer [13]. The study noted human factors related accidents in Boeing and Airbus aircraft had occurred significantly less than expected relative to the number of operational aircraft. Another relevant study compared the safety performance of Boeing and Airbus between 1990 to 1998 [14]. The study showed that in a direct comparison, correcting for fleet size, the probability of an accident occurring was greater for Boeing than for Airbus. A related study looked at accidents only involving avionic systems [15]. This noted that while the Boeing 737 was more frequently associated with these accidents, most of these were still attributed to human factors, highlighting the critical role of the end user.

The aim of this research is to examine the International Civil Aviation Organization (ICAO) safety occurrence data from 2008 to 2019, comparing the Boeing 737 and Airbus A32x families of aircraft. The underlying research question is "are there any statistically significant differences in the safety occurrences of the core narrow-body single-aisle aircraft of Boeing and Airbus?" To achieve this goal, a brief background will provide details of the families of aircraft, followed by details of the methodology. The data will be presented in four parts. The first three will look at aspects of 1) all safety occurrences (accidents and incidents), 2) all accidents, and 3) fatal accidents. It should be noted that each consecutive set is a subset of the proceeding set. The fourth aspect will compare annual fatal and hull loss accidents per million departures.

## II. BACKGROUND

### A. Boeing 737

The Boeing 737 family has a long history, beginning in 1964 [16]. The first generation, including the 737-100 and 737-200, began commercial operations in 1967. The basic Boeing 737 is illustrated in Fig. 1. The second generation, including the 300, 400, and 500, began commercial operations in 1984, 1988, and 1989, respectively. It should be noted that the typical two cabin

Color versions of one or more of the figures in this article are available online at http://ieeexplore.ieee.org

seating configuration of the 100 and 200 had passenger capacities of 85 and 102, respectively; for the 300, 400, and 500, these increased to 126, 147, and 110, respectively. That is, the 300 and 400 offered increased capacity with greater performance, while the 500 was a "modern" replacement of the 200 (modern in terms of incorporating recent cockpit and engine improvements).

The third generation of the Boeing 737 family commenced operations in 1997 [16]. Each of the aircraft in this generation effectively replaced a previous iteration. The 600 replaced the 500, the 700 replaced the 300, and the 800 replaced the 400, all now in a logical sequence reflecting their size. The aircraft was stretched further for the 900, which has a seating capacity of up to 220, making it a replacement for the 757-200. The fourth generation, the Max series, represents a similar refresh to the 737 range, with the next generation of engine upgrades, along with aerodynamic improvements, increasing performance and efficiency. The Max 7, 8, and 9, replace the 700, 800, and 900, respectively. A Max 10 has also been developed, again offering an increased capacity. The Boeing 737 Max entered service in 2017.

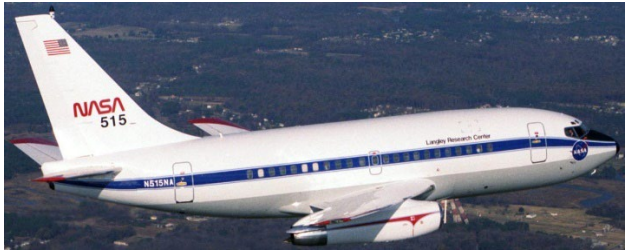

**Fig. 1.** The 1st Boeing 737-100 produced in 1967 became the last delivered in 1973, to NASA.

*B. Airbus A32x*

The term A32x applies to the Airbus family of aircraft that are similar to the Boeing 737 family. Specifically, this includes the A318, A319, A320, and A321 [17]. The series started with the A320 aircraft [18], which first entered operation in 1988 [19], and was similar in capacity and performance to the Boeing 737-400 at the time (and later the 800). This is relevant to the Max accidents, as the Boeing 737 had started as a much smaller aircraft, and the maneuvering characteristics augmentation system (MCAS) was needed due to the larger engines [8]. For Airbus to compete with the smaller variants of the Boeing family, the A320 was shrunk to the A319 (beginning operational service in 1996), which competed with the 737-300 (and later the 700). The design was shrunk further to give the A318 (entering service in 2003), which competed with the Boeing 737-600. Only one stretch redesign was needed to give the higher capacity A321, which commenced operations in 1994 [20]. The interesting point here is that the A321 predates its Boeing equivalent, the 737-900.

As with the Boeing aircraft, there have been upgrades and improvements to the Airbus product line. Of note is the enhanced variants, which was followed by the new engine option (NEO). NEO is the Airbus program equivalent to the Boeing Max program. The A321XLR is the most recent version announced [21]; the A319NEO is shown in Fig. 2.

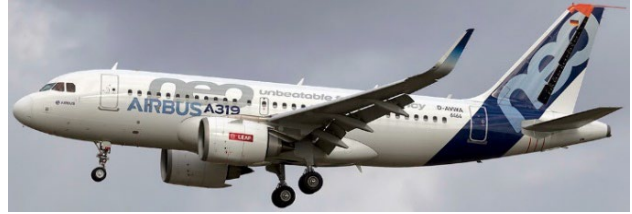

**Fig. 2.** The Airbus A319NEO.

*C. Summary*

The details of the previous two sections are summarized below in Fig. 3. This includes a timeline and a table indicating the cabin capacity and range of the various aircraft.

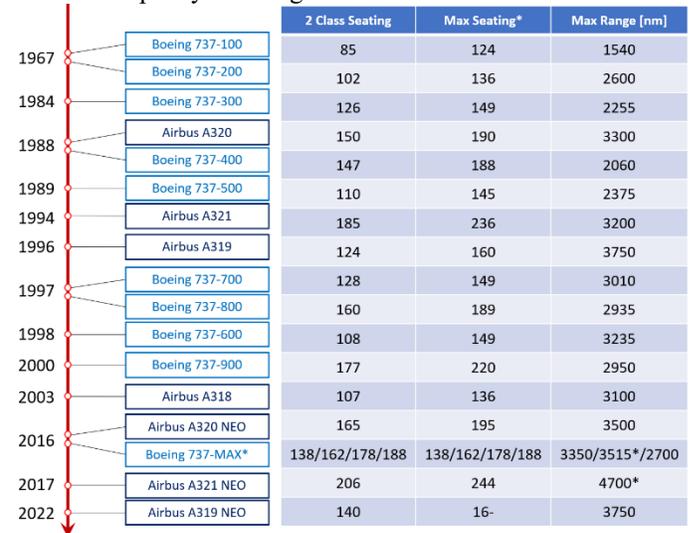

| Year | Aircraft | 2 Class Seating | Max Seating* | Max Range [nm] |
|---|---|---|---|---|
| 1967 | Boeing 737-100 | 85 | 124 | 1540 |
|  | Boeing 737-200 | 102 | 136 | 2600 |
| 1984 | Boeing 737-300 | 126 | 149 | 2255 |
| 1988 | Airbus A320 | 150 | 190 | 3300 |
|  | Boeing 737-400 | 147 | 188 | 2060 |
| 1989 | Boeing 737-500 | 110 | 145 | 2375 |
| 1994 | Airbus A321 | 185 | 236 | 3200 |
| 1996 | Airbus A319 | 124 | 160 | 3750 |
| 1997 | Boeing 737-700 | 128 | 149 | 3010 |
|  | Boeing 737-800 | 160 | 189 | 2935 |
| 1998 | Boeing 737-600 | 108 | 149 | 3235 |
| 2000 | Boeing 737-900 | 177 | 220 | 2950 |
| 2003 | Airbus A318 | 107 | 136 | 3100 |
|  | Airbus A320 NEO | 165 | 195 | 3500 |
| 2016 | Boeing 737-MAX* | 138/162/178/188 | 138/162/178/188 | 3350/3515*/2700 |
| 2017 | Airbus A321 NEO | 206 | 244 | 4700* |
| 2022 | Airbus A319 NEO | 140 | 16- | 3750 |

**Fig. 3.** Timeline and comparison for relevant Boeing and Airbus aircraft. Notes (*): Maximum seating is as certified, A321NEO range is XLR, MAX* is 7, 8, 9, & 10 in order, with the 8 and 9 having the same range.

III. METHODOLOGY

The research design implemented in this work follows that of related previous research. The fundamental aspects of the research design including the methodology has been previously described in detail [22]. Given a set of nominal data, this is described by the cross-sectional component, the two aircraft manufacturers, and several key variables. Each of these variables, occurrence type, occurrence category, phase of flight, injury level, and number of fatalities, has two or more nominal values. The distribution of these values in terms of their counts across a variable is then compared, either to some expected distribution (a goodness of fit test) or between cross-sectional elements (a test of independence). Where the samples between Airbus and Boeing are tested for their independence, the appropriate Chi Squared test was utilized [23]. This involved counting the relevant number of occurrences with each value, in each variable, for each manufacturer. The distribution of each variable was then compared between the manufacturers. It should be noted that Boeing has higher counts, and as such, data presentation utilizes the relative proportion, noting the total count of each, such that the distribution of proportions can be directly compared visually. Where the relative proportions between Airbus and Boeing occurrences were compared to an expected distribution based on the total of all Airbus and Boeing occurrences, a goodness of fit test was utilized [24]. This was used for the number of fatalities (the count), where Airbus and Boeing became the nominal values. Two expected distributions were tested; an even split relative to the total number of 1) occurrences, and 2) accidents. This was selected as Boeing being utilized more in terms of operations, will unfairly have a higher number of fatalities. The data presentation here utilized the observed count relative to the expected proportion, because each manufacturer is being compared to itself. For very small sample sizes (in cases violating the requirements to use a chi squared test), a Fisher Exact test was needed [25]. The data utilized in this work has been used previously for similar studies [26], and is publicly available from ICAO [27].

## IV. RESULTS AND ANALYSIS

### A. All Safety Occurrences

Fig. 4 shows the breakdown of all the safety occurrences coded by type of occurrence (accidents, serious incidents, or incidents). The Chi Squared test shows a statistically significant difference between the proportions of Airbus (a) and Boeing (b) coded by type of occurrence (a = 2,826, b = 2,593, $\chi^2$ = 8.37, p = 0.015). The greatest contribution to the result is the proportion of accidents, which is 10% greater than expected for Boeing, and 9% less than expected for Airbus.

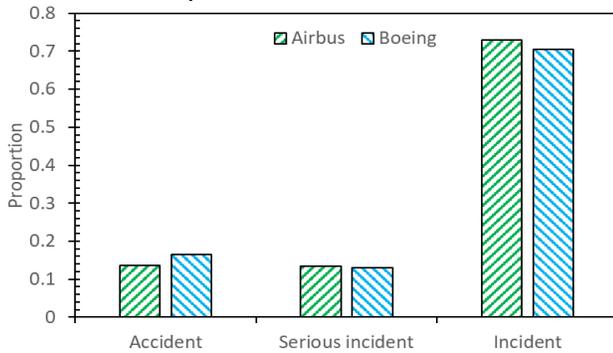

**Fig. 4.** Proportions of all safety occurrences coded by the type of occurrence. Total counts: Airbus = 2,826, Boeing = 2,593.

Fig. 5 shows the breakdown of all the safety occurrences coded in terms of the ICAO occurrence categories [28]. The Chi Squared test shows a statistically significant difference between the proportions of Airbus (a) and Boeing (b) in terms of occurrence categories (a = 843, b = 943, $\chi^2$ = 61.2, p < 0.001). Two specific categories have significant counts that contribute to this result. The first is midair collisions (MAC), which are 23% greater than expected for Airbus, and 21% less than expected for Boeing. The second is runway excursions (RE), which are 36% less than expected for Airbus and 32% more than expected for Boeing. While MACs are serious, it should be noted that an event is still recorded as a midair collision even if corrective action is taken to prevent it; however, the event is just an incident and not an accident (there is no injury, damage, loss of life or airframe). The excess MAC for Airbus could be attributed to airspace issues in which these aircraft are operated. Similarly, the excess of RE for Boeing could be attributed to aerodromes issues where they are operated. Further research would be needed to assess the veracity of these hypotheses.

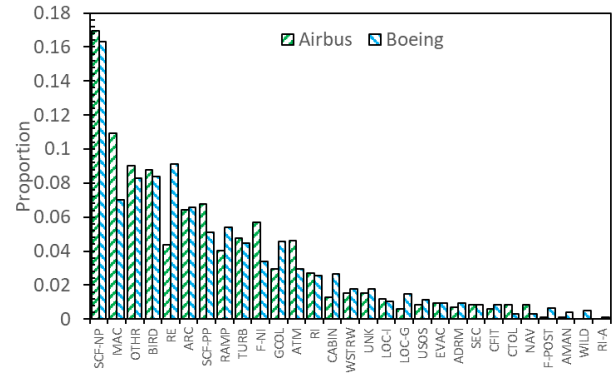

**Fig. 5.** Proportions of all safety occurrences coded by ICAO occurrence category. Total counts: Airbus = 843, Boeing = 943.

Fig. 6 shows the proportions of safety occurrences coded in terms of phase of flight. The Chi Squared test shows an almost statistically significant difference between the proportions of Airbus (a) and Boeing (b) coded by phase of flight (a = 811, b = 874, $\chi^2$ = 14.0, p = 0.052). As such, a follow up test was conducted, noting the spike in proportions for Boeing at landing. That is, landing was compared to all other phases of flight, collectively, testing for independence between Airbus and Boeing. This result was statistically significant ($\chi^2$ = 3.95, p = 0.047). Combined, there appears to be some evidence to support the finding that Boeing has more safety occurrence than expected during landing by 9%. It should be noted that Boeing had 24 more RE events than expected and 30 more occurrences at landing than expected. As such, it is possible the increased number of Boeing landing events is likely due to the aerodromes where the aircraft are operated, as previously noted.

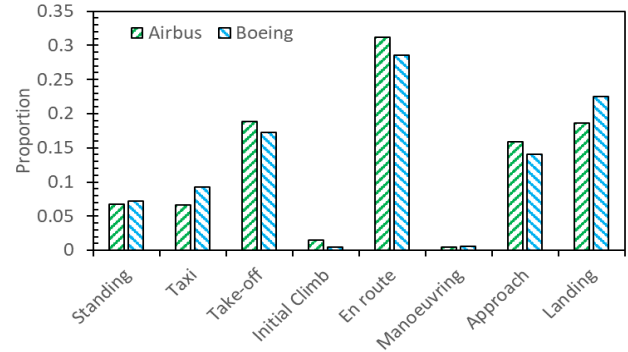

**Fig. 6.** Proportions of all safety occurrences coded by phase of flight. Total counts: Airbus = 811, Boeing = 874.

Fig. 7 shows the proportions of all safety occurrences coded in terms of the injury level. The Chi Squared test is not statistically significant (a = 853, b = 985, $\chi^2$ = 2.83, p = 0.42). Looking at fatalities (shown in Fig. 8), relative to the total number of safety occurrences, the number of observed fatalities in Boeing occurrences is 28% more than expected, while for Airbus they are 26% less than expected. This result is statistically significant (a = 773, b = 1,251, $\chi^2$ = 144, p < 0.001).

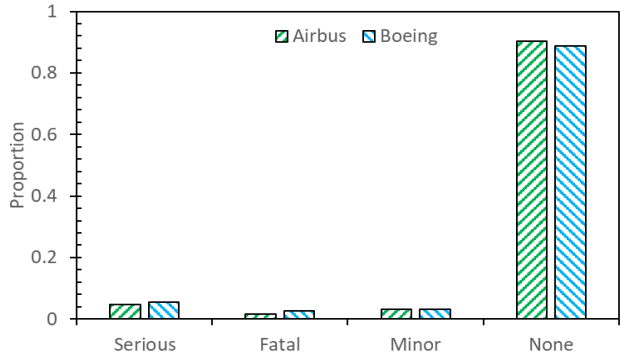

**Fig. 7.** Proportions of all safety occurrences coded by injury level. Total counts: Airbus = 853, Boeing = 985.

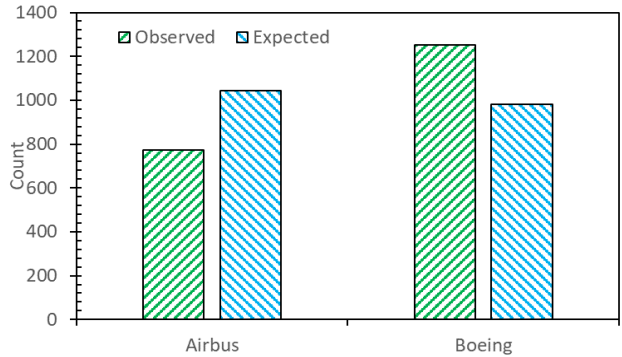

**Fig. 8.** Number of fatalities for Airbus and Boeing (expected based on the number of occurrences).

*B. Accidents*

Fig. 9 shows the breakdown of accidents coded in terms of ICAO occurrence categories. When looking at only accidents rather than all safety occurrences, the observed difference in distribution is not statistically significant (a = 182, b = 283, $\chi^2$ = 10.8, p = 0.29). The proportions for abnormal runway contact and turbulence are higher for Airbus (a) while the system component failure non-powerplant and runway excursions are higher for Boeing (b); although, not statistically significantly different.

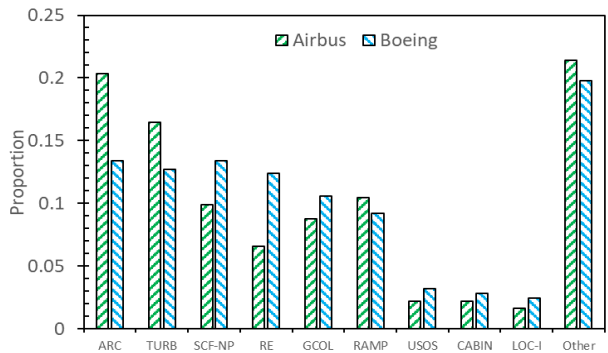

**Fig. 9.** Proportions of accidents coded by occurrence category. Total counts: Airbus = 182, Boeing = 283.

Fig. 10 shows the proportions of accidents coded in terms of phase of flight. The Chi Squared test shows the difference in proportions between Airbus (a) and Boeing (b) are not statistically significantly different (a = 214, b = 263, $\chi^2$ = 7.2, p = 0.41). The follow up test comparing landing to all other phases of flight, between Airbus and Boeing was also not statistically significant ($\chi^2$ = 0.05, p = 0.82).

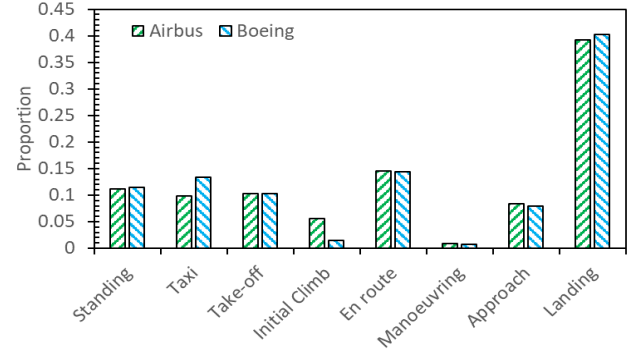

**Fig. 10.** Proportions of accidents coded by phase of flight. Total counts: Airbus = 214, Boeing = 263.

Fig. 11 shows the proportions of all the safety occurrences coded in terms of the injury level. The Chi Squared test is not statistically significant (a = 853, b = 985, $\chi^2$ = 2.83, p = 0.42). In terms of fatalities relative to the number of accidents (shown in Fig. 12), the number of observed fatalities in Boeing accidents is 18% more than expected, while for Airbus they are 20% less than expected. This result is statistically significant (a = 771, b = 1,249, $\chi^2$ = 70, p < 0.001).

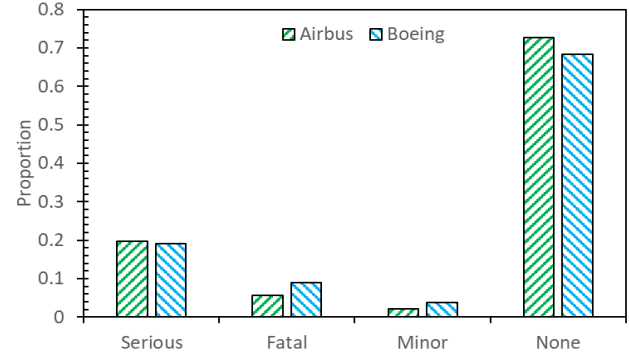

**Fig. 11.** Proportions of accidents coded by injury level. Total counts: Airbus = 853, Boeing = 985.

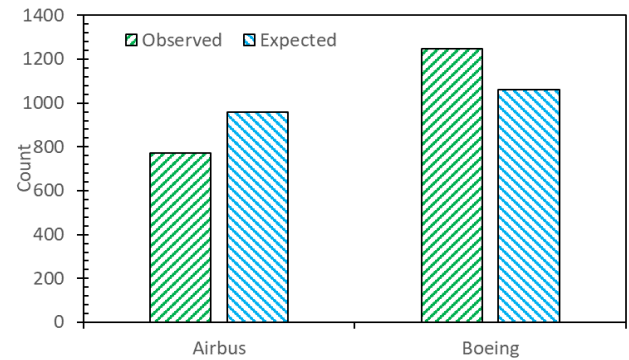

**Fig. 12.** Number of fatalities for Airbus and Boeing accidents (expected based on the number of accidents).

*C. Fatal Accidents*

Finally, we can consider just fatal accidents. Fig. 13 shows the distribution of the Airbus (a) and Boeing (b) fatal accidents coded by ICAO occurrence categories. Due to the small number of events, the broad category of runway safety has been used to group all relevant categories. Using a Fisher exact test, the result is statistically significant (p = 0.032). Airbus has three security and three ramp fatal accidents, while Boeing has none. Boeing has three fire post impact and two system component failure powerplant events, while Airbus has none.

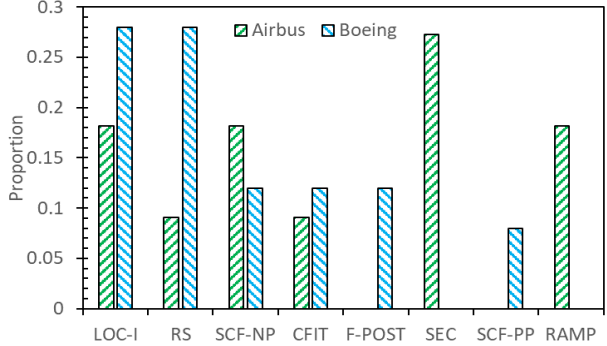

**Fig. 13.** Proportions of fatal accidents coded by occurrence category. Total counts: Airbus = 11, Boeing = 25.

The phase of flight for the fatal accidents was also analyzed. The corresponding Fisher exact test was not significant (p = 0.26). Finally, as these are all fatal accidents, the injury level for all of them is fatal, and hence no comparison is possible.

*D. Accident Rate*

The final aspect to consider is the accident rate, both the hull loss rate and the fatal hull loss rate, relative to the number of departures. This data is provided in the "Boeing Annual Summary of Commercial Jet Airplane Accidents", available online [29]. Data was utilized from 2006 to 2020. Prior to 2006 the reports only provided the rate for hull losses, and not fatal hull losses. Fig. 14 shows the hull loss rate per million departures for the Boeing 737 family and the A32x family of aircraft, while Fig. 15 shows the fatal hull loss rate per million departures. As can be seen in both plots, the relative reduction in both types of accidents was greater for Airbus relative to Boeing. Both plots have a similar relative vertical span, relative to zero, which highlights that the 2006 values of Boeing are greater; however, this fact can be explained by the greater number of Boeing aircraft in operation, performing more departures. In 2006, Boeing 737 aircraft performed 4.2 times as many departures compared to Airbus A32x aircraft. By 2019, this had reduced to a factor of 1.7. That is, the number of departures performed by Airbus A32x aircraft has grown at a greater rate than that of Boeing 737 aircraft. However, the rate is a relative measure, to the number of departures. As such, any statistically significant difference in the change of rate is noteworthy. To enable a comparison, both the, CAGR, compound annual growth rate (effectively the average annual interest rate since the start) and the relative annual change (this year's value subtract last year's value, divided by this year's value) were computed, for both the hull loss rate and fatal hull loss rate. Both rates were calculated, as CAGR is sensitive to offsets. That is, the effect of a step change (such as a linear trend stepping from $y = x$ to $y = x +1$ after $x = 10$) influences every CAGR calculation after the step, while for the relative rate, it only influences the year of the change.

The summary data is shown in Table 1. Using a two-sample t-test with one tail, all 4 tests show that the annual reduction in the accident rates for Airbus are better than that for Boeing. That is, the rate of accidents per departure of Airbus is reducing faster than for Boeing. This is explained by the much more significant growth in operations with Airbus aircraft, without a corresponding increase in hull losses, fatal or otherwise. As expected, the p-values for the CAGR are smaller.

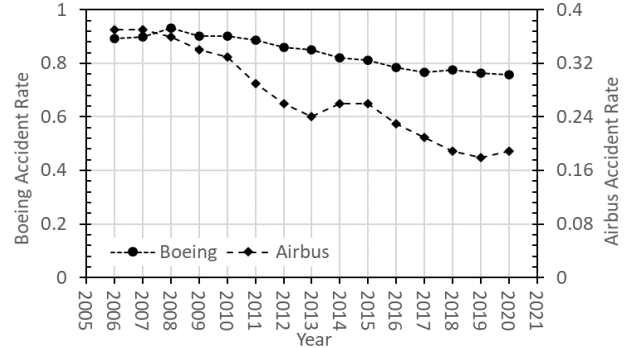

**Fig. 14.** Annual hull loss rate per million departures.

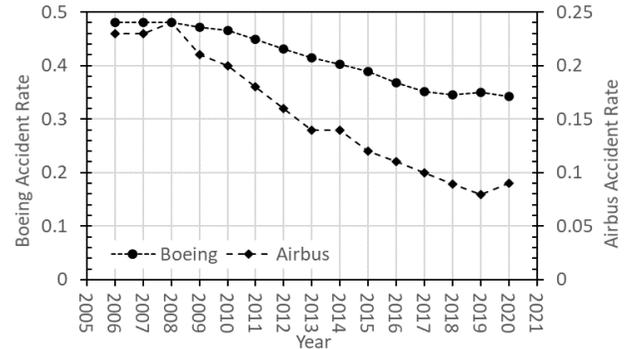

**Fig. 15.** Annual fatal hull loss rate per million departures.

TABLE I
MEAN (μ), STANDARD DEVIATIONS (σ), AND ONE TAILED TWO SAMPLE T-TEST P VALUES FOR ANNUAL REDUCTION IN RATES OF HULL LOSS (HL) AND FATAL HULL LOSS (F-HL) RATES, IN TERMS OF RELATIVE ANNUAL REDUCTION (Δr/r) AND CAGR. *KOLMOGOROV-SMIRNOV TEST FOR NORMALITY CRITICAL VALUE 0.26.

| Metric | Rate | Manu | μ | σ | p | KS* |
|---|---|---|---|---|---|---|
| Δr/r | HL | Boeing | -0.012 | 0.020 | 0.023 | 0.10 |
| | | Airbus | -0.051 | 0.067 | | |
| | F-HL | Boeing | -0.025 | 0.020 | 0.021 | 0.16 |
| | | Airbus | -0.073 | 0.081 | | |
| CAGR | HL | Boeing | -0.004 | 0.008 | <0.001 | 0.17 |
| | | Airbus | -0.036 | 0.016 | | |
| | F-HL | Boeing | -0.016 | 0.009 | <0.001 | 0.19 |
| | | Airbus | -0.046 | 0.027 | | |

V. DISCUSSION

Several of the metrics considered in this study, in total across these two families of narrowbody aircraft, question the traditional belief that Boeing aircraft are "safer" than their Airbus counterparts. This does not at all suggest that Boeing aircraft are not safe. This is a comparison between the two safest aircraft manufacturers [13]. It is an examination of the two safest to see if there are any observable statistically significant differences between them. The results shown in Fig. 14 and 15 clearly highlight that both have improved in safety when looking at the metric of hull loss accident rate per million departures and fatal hull loss accident rate per million departures. However, the results indicate that accident rates in Airbus aircraft have reduced at a greater rate, and in terms of severity of occurrence type and number of fatalities, Airbus is below expected, while Boeing is above expected, only when compared together. Relative to all other manufactures, previous studies have shown that both Boeing and Airbus perform better [13]. The conclusion that suggests Airbus is performing better than Boeing in terms of safety mirrors previous studies [14, 15].

A reader may ask the question, what would happen if the "anomalous" data points of the Boeing 737MAX were removed? The findings for fatalities relative to all safety occurrences (Fig. 8) remains statistically significant in Airbus' favor, while fatalities relative to accidents (Fig. 13) becomes statistically insignificant, still in Airbus' favor. So, at best, one could say that on some metrics Airbus and Boeing are level, while in other Airbus is outperforming Boeing. However, the case to exclude these as so called "black swan events" [30], is not justifiable. In fact, while the total of this data set represents the reported occurrences for Boeing and Airbus end users, the Boeing 737MAX occurrences are events which were more associated with the manufacturer than the end user. To yield unbiased comparison between the Airbus and Boeing data requires all of it to be considered. Furthermore, prior to 2018, Boeing 737 aircraft were experiencing 1.55 fatal hull loss accidents per year, while up to 2020 this only increased to 1.65 fatal hull loss accidents per year. This variation is well within a traditional uncertainty. As such, claiming that the two Boeing 737MAX accidents are "black swan" events, and hence should not be factored in is not justifiable scientifically. In fact, this would constitute "cherry picking", a common technique of science misinformation [31].

The important context of these numbers is the population. To be clear, the total number of Boeing aircraft in operation during all years of this study was significantly greater than the number of Airbus aircraft. As noted, Boeing 737 aircraft performed 4.2 times as many departures compared to Airbus A32x aircraft in 2006, which reduced to 1.7 times by 2019. The purpose of using the rate per million departures is to remove this population skew. Just looking at the rates in 2006 to 2008 shows that Airbus and Boeing were on par. The average age of the Boeing fleet might play a role in this. However, the age of the fleet is an objective feature of that fleet. Furthermore, it should be noted that for technical issues, such as mechanical fatigue, issues are associated more with a period up to the first D check than they are for older airframes [22].

A core assumption in the work is that the ICAO occurrence data set as a sample is representative of the population. There is no reason to doubt the validity of this assumption. This is supported by the fact that Annex 13 to the Chicago Convention requires that Contracting States report to ICAO information on all aircraft accidents which involve aircraft of a maximum certified take-off mass of over 2,250 kg [32]. Further to this, all serious incidents involving aircraft with a maximum certified take-off mass of over 5,700kg are also collected.

The reliability of the data, in terms of completeness and correctness, as available from the ICAO database is assumed. There is no independent way to verify this, and a study by the European Spreadsheet Risks Interest Group reported that less than one in ten spreadsheets are error free [33]. Given most errors are computational errors, a database as presented is taken as complete and correct. Furthermore, there is no reason to suspect that errors would favor Boeing or Airbus over the other.

The results presented here in terms of statistical inferences are the result of the aggregate of the data over the period of 2008 to 2019. Those statistical inferences can only describe if observed differences in the proportions between the samples are statistically significantly different. Why any confirmed independence between Airbus and Boeing categories exists cannot be determined from this purely quantitative data. While this is an accepted limitation of the methodology, it does not negate that the observed differences are statistically significant compared to random chance alone.

Future work is proposed to conduct a collective case study on the statistically significant findings of this work. The aim of this future work will be to understand why Airbus appears to be involved in more MAC occurrences and why Boeing appears to be involved in more RE occurrences than expected. This would simultaneously address the potential spike in occurrences at landing for Boeing relative to Airbus.

## VI. CONCLUSION

The analysis in this work highlighted several categories with statistically significant differences. When looking at all safety occurrences, the occurrence type, occurrence category, and number of fatalities were significant. For occurrence type the proportion of accidents was 10% greater than expected for Boeing and 9% less than expected for Airbus. For occurrence category, midair collisions were 23% greater than expected for Airbus, while runway excursions were 32% more than expected for Boeing. For number of fatalities, Boeing had 28% more than expected, while Airbus had 26% less than expected. Looking at just accidents, only the number of fatalities was statistically significantly different; specifically, Boeing accidents had 18% more fatalities than expected, while Airbus accidents had 20% less fatalities than expected. In terms of the hull loss and fatal hull loss rates per million departures, Airbus rates have reduced more than Boeing over the past 15 years, at a statistically significant level.

While the saying "if it ain't Boeing I ain't going" has existed since the time of the 707 [34], it became significant in response to the perceived issue with Airbus and fly-by-wire [35]. This research highlights that the sentiment is still true in relation to any other manufacturer, but not relative to Airbus.

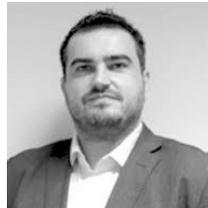

**Graham Wild** (Member, IEEE) was born in Rotherham, England in 1981. He received his Bachelor of Science degree in Physics and Mathematics as well as his Bachelor of Science Honors degree in Physics from Edith Cowan University, Joondalup, WA, Australia in 2004 and 2005, respectively, his Graduate Certificate in Research Commercialization from Queensland University of Technology, Brisbane, QLD, Australia in 2008, his Master of Science and Technology degree in Photonics and Optoelectronics from the University of New South Wales, Sydney, NSW, Australia in 2008, and his PhD in Engineering from Edith Cowan University, in 2010.

He is currently the aeronautical engineering program coordinator and a senior lecturer in Aviation Technology the University of New South Wales, in Canberra, Australia. From 2012 to 2019, he was a senior lecturer in Aerospace Engineering and Aviation at RMIT University, in Melbourne, Australia. From 2011 to 2012 he was a lecturer in Aircraft Systems at Edith Cowan University, in Joondalup, Australia. In 2010 he was a post-doctoral research fellow with the Optical Research Laboratory at Edith Cowan University, in Joondalup, Australia. He has completed two research internships with the Australian CSIRO Industrial and Telecommunications Physics Division in 2003 to 2004 (with the Electric Machines Group), and 2004 to 2005 (with the Intelligent Systems Group), in Sydney, Australia. His current area of research is focused on intelligent systems, AI, data and analytics, and advanced technology in aviation and aerospace, for safety and sustainability.

Dr Wild is a member of the IEEE, SPIE, and AIAA.